\documentclass[a4paper,twoside]{article}

\usepackage{balance}
\usepackage{epsfig}
\usepackage{subcaption}
\usepackage{calc}
\usepackage{amssymb}
\usepackage{amstext}
\usepackage{color}
\usepackage{amsmath}
\usepackage{amsthm}
\usepackage{multicol}
\usepackage{pslatex}
\usepackage{apalike}
\usepackage{algorithm2e}
\usepackage[bottom]{footmisc}
\usepackage{SCITEPRESS}     

\begin{document}

\title{The Perceived Learning Behaviors and Assessment Techniques of First-Year Students in Computer Science: An Empirical Study}

\author{\authorname{Manuela Petrescu\sup{1}\orcidAuthor{0000-0002-9537-1466}, Tudor Dan Mihoc\sup{1}\orcidAuthor{0000-0003-2693-1148}}
\affiliation{\sup{1}Department of Computer Science, Babe\c s Bolyai University, Cluj-Napoca, Romania}
\email{manuela.petrescu@ubbcluj.ro, tudor.mihoc@ubbcluj.ro}}


\abstract{The objective of our study is to ascertain the present learning behaviors, driving forces, and assessment techniques as perceived by first-year students, and to examine them through the lens of the most recent developments (pandemic, shift to remote instruction, return to in-person instruction).  Educators and educational institutions can create a more accommodating learning environment that takes into account the varied needs and preferences of students by recognizing and implementing these findings, which will ultimately improve the quality of education as a whole.
Students believe that in-person instruction is the most effective way to learn, with exercise-based learning, group instruction, and pair programming. Our research indicates that, for evaluation methods, there is a preference for practical and written examinations. Our findings also underscore the importance of incorporating real-world scenarios, encouraging interactive learning approaches, and creating engaging educational environments.}

\onecolumn \maketitle \normalsize \setcounter{footnote}{0} \vfill

\section{\uppercase{Introduction}}
\label{sec:introduction}

Education in the present day faces a number of novel challenges. These problems originate either from global unfortunate events, such as the pandemic, or from mandated educational activities, such as distance e-learning, as a result of these events \cite{gutierrez2023learning}. These difficulties have had a substantial impact on the quality of education and students' ability to learn. As a result, students may struggle to keep track of their schoolwork and fall behind; their academic performance may suffer as a result, as will their motivation. This will eventually jeopardize their ability to achieve their academic objectives and realize their full potential. Therefore, it is critical that teachers provide enough support and resources to help students overcome these obstacles and stay on track with their studies.

The COVID-19 pandemic forced both teachers and students to switch from traditional in-person instruction to online learning environments \cite{Lemay2021TransitionTO}.

Although most campuses have faculty members with training to guarantee the quality of the content, few have created strategies for this type of interaction with students. Some suggestions to improve student participation and engagement in online learning environments are provided in \cite{neuwirth2021reimagining}.  They proposed a framework that academic departments might use to create policies that cover best practices and recommendations for post-pandemic synchronous and asynchronous virtual classrooms. 

Any attempt to solve these problems is hampered by the lack of knowledge about the preferences and study habits of the students after the quarantine period. We attempt to close this knowledge gap in this study by finding out what students think is the best approach to carrying out the teaching and learning process. 

\section{\uppercase{Literature Review}}

Several strategies have been proposed to increase the efficacy of instruction \cite{liu2022increasing,Seikkula}, with many focusing on computer science teaching methods \cite{salas2017teaching,Marcus2020}. In addition, some offer resources to computer science instructors \cite{Hazzan2020}, and suggest technical professional development and collaboration tactics for teachers \cite{Wang21}.

Several research papers have been published on the topic of online student motivation \cite{baquerizo2020motivacion}, online course assessments \cite{v2018}, and focus of elements such as teacher collaboration strategies and technical professional development \cite{Wang21}.

Online classes have the benefit of allowing students to learn at their own pace using text or video resources, which can facilitate learning. Before lectures, students can access the materials, which allows them to ask more questions and complete more difficult assignments. \cite{s2020}.

The student dropout rate is a problem educators face on a regular basis. The percentage of students enrolled who do not graduate with a degree or dropout varied significantly in European countries, from a low 14,4\% in the UK, to approximately 29\% in Germany, according to \cite{heuble2017}. The highest value, according to EU reports \cite{UEDropOut} is 60,6\% in France.  
Online or hybrid learning environments \cite{de2022characteristics} were also considered as a response to dropout rates in traditional classroom settings. There are other factors that affect the dropout rate, including academic achievement, starting age, and parental educational background \cite{Araque}. 

Fear of failure can also be a factor that slows academic development. \cite{Martino1993AGM} offers some goal-setting strategies that can help young students avoid becoming \textit{failure-accepting} students.

There is a lot of literature related to mechanisms that can be employed by people when faced with challenging circumstances that carry a high risk of failure and possible harm to their self-esteem \cite{Norem1986DefensivePH}.  In the literature, as guideline for educators, the main approach to this matter is mostly by creating a supportive and nurturing environment, provide educational treatment, and use autonomy-supportive teaching styles.

\section{\uppercase{Study Setup}}

\hspace{\parindent} \textbf{Scope:} We aim to determine the current learning habits, motivations, and evaluation methods in the freshmen students opinion and analyze them from the perspective of the changes in the last period (COVID, transition to online teaching, transition back to face-to-face teaching). 

The COVID-19 pandemic's restrictions on this generation of students also had a direct impact on their initial years of high school study: social isolation, a sudden switch to online instruction for a year, and after to a return to in-person instruction from teachers.

We were interested in students' perspective on a stage in life when, after graduating high school and enrolling at the university , they had little time to adjust to the new academic environment.

We aim with this up-to-date data to fill in the gaps in our understanding of the students. 

The research questions are:
\begin{itemize}
    \item [-] What drives students' primary motivation to acquire knowledge?
    \item [-] Which evaluation method is considered by students to be the most equitable, and why?
    \item [-] What do students think are the most effective ways to learn?
\end{itemize}

The authors formulated these research questions after defining the study objective. 

\textbf{Setup:}
In order to explore the answers to our research questions, we designed a survey addressed to young students. We sent the survey link to students in the eighth week of the first semester, which were randomly allocated to one of the authors of this research in the Computer Science Architecture. 

We outline the purpose of collecting their input and the intended use of the gathered data in an effort to boost engagement. A brief speech about our research and expressing our appreciation for their time in responding served as the sole source of motivation. We also noted the anonymity of the survey.  

\textbf{Participants:}
A group of 43 students enrolled in the first year of computer science in the English line of study participated in the survey. The initial set of participants consisted of 67 students, of whom 43 decided to participate. The selection of the 67 students that were required to participate was aleatory; the students were randomly selected from the students assigned to classes with one of the authors of this study. 

There was no discrimination between the students; all received the survey link. Participation was optional and we did not use any reward mechanism for participation, except for the motivational speech mentioned above. We obtained 64.17\% responses from the initial set of participants. 

\textbf{Study design:}
After analyzing and deciding on the research questions, we designed the survey. The process of elaborating the questions was made up of a series of steps: each author proposed some questions that formed the first draft and we discussed, analyzed, and modified the questions until we reached consensus. 

We used open questions because they provide more information about the students' thoughts and can better reflect their perceptions of reality. The questions were asked in the student's native language, even if the participants were enrolled in the English line to obtain more descriptive and complex answers. We asked positive and negative questions to avoid bias, and we asked the students to motivate their responses rather than specify a reason or a learning method. We provide the learning methods examples only in the last question as we did not want to influence previous responses. The list of survey questions is shown in Table \ref{tab:questions}.

\begin{table*}[t]
    \centering
    \begin{tabular}{|l|l|}
    \hline
        \ \ Q1 \ \ & What are the factors that motivate you to learn? \\
        \hline
        \ \  Q2 \ \ & What are the factors that discourage learning a topic or a course? \\
        \hline
        \ \ Q3 \ \ &  What do you consider to be the worst form of evaluation and why? \\ 
        \hline
         \ \ Q4 \ \ &  What do you consider to be the best form of evaluation and why?\\
         \hline
         \ \ Q5 \ \ &  Which are in your opinion the best teaching methods?   \\
          \hline
        \ \ Q6 \ \ &  Which teaching methods worked best for you? What do you use? \\&  (online courses, face-to-face teaching, pair programming) \\
         \hline
    \end{tabular}
    \caption{Survey Questions}
    \label{tab:questions}
\end{table*}

\subsection{Methodology}
In the eighth week of the second semester, we conducted an online survey using both accountable and open questions. Accountable questions facilitate working with and interpreting some facts, and open questions offer a further level of insight. To evaluate and interpret responses to open questions, we used quantitative techniques, particularly questionnaire surveys, according to the guidelines established by the empirical community \cite{ACM}. For the interpretation of the text, we apply the definition of theme analysis found in \cite{Braun19}. These techniques have already been applied in other studies related to computer science \cite{petrescu2023perspective,enase21,PETRESCU20231028}.

The following steps constitute the text analysis methodology.

(1) In order to identify the keys in the text, two researchers performed an independent analysis.

(2) Using techniques that include generalization, removal, and reassignment of significant items with low prevalence to similar themes or classes, we classified important items based on shared themes or classes.

(3) In the last phase, all authors review the process with careful consideration in order to assess the degree of trust in the methodology. Examining and debating numerous topics, explanations and corroborating data related to the categorization procedure was part of this process. 

We calculated the frequency of the keywords used in the answers. Some student submissions featured only one or two ideas, while others provided up to four arguments to support their choice of research topics. Thus, an answer could include additional things or significant words. We used the calculated prevalence of several key items, classified them, and compared them with the total number of responses received. As a result, the total percentages will exceed 100\%.

\textbf{Data collection:}

Several steps were taken in the process of developing survey questions. Initially, we developed the study objectives and determined the scope of the article. Each author supplied a list of survey questions based on these. We studied and debated the proposed collection of questions before deciding which questions would be selected for the survey.

The questions were translated from English into the students' native language, allowing them to respond in their preferred language. We used this strategy to promote student participation. We employed automatic language translation technologies to translate them into English before having the authors double-check the translation.

To increase the number of responses, we left the survey open for two weeks. We distributed the survey link to the student group so we could gather responses anonymously, allowing students to respond when and how they chose. 

The collected answers were pre-processed prior to analysis, eliminating some special characters, empty lines, and spaces.

\subsection{Q1 - What drives students' primary motivation to acquire knowledge?}
Motivation is a key component in the learning process, it can be positively or negatively influenced, depending on a sum of factors. We wanted to explore which are these factors for first-year students in computer science, as in this period, the are still deeply involved in the learning process but are mature enough to appreciate and take into consideration a large broad range of factors. We asked them to describe what factors motivate them to learn a course and what factors discourage them.\\
\textbf{What are the factors that motivate students to learn the content of the course?}

\begin{figure}[htbp!]
\centering
\includegraphics[width=0.45\textwidth,]{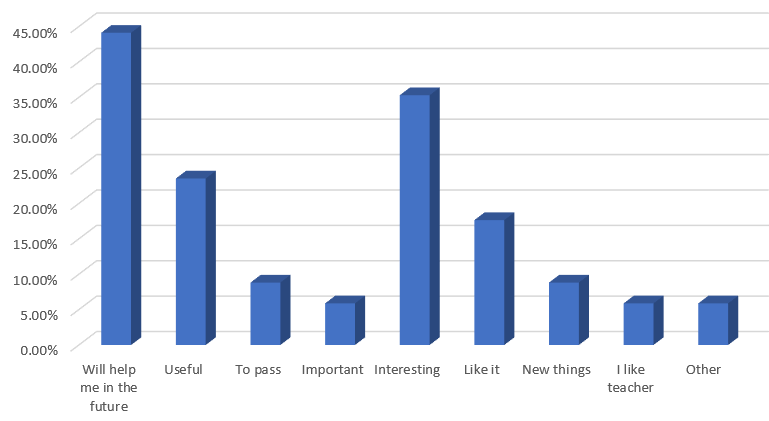}
\caption{Factors that motivate students to learn} \label{fig:ReasonsToLearn}
\end{figure}

We grouped the factors mentioned by students in two major classes, one set of factors are related to economical perspectives and the second set of factors related to personal reasons:
\begin{itemize}
    \item Economic: Will help me in the future, Useful, To pass (the exam), Important 
    \item Personal: Interesting, I like (the topics), I like the teacher 
\end{itemize}
A set of factors such as \textit{''To learn how different computer programs work''} have a small prevalence, so we decided to place them in the class \textit{''Other''}. The percentages of motivational factors, grouped by classes, mentioned by the students are presented in Figure \ref{fig:ReasonsToLearn}.

Economic factors seem to have a higher prevalence compared to personal / emotional factors. The responses of many students reflected this aspect: \textit{''If the things I learn in the course are up-to-date and will help me in the future job'',''If the course is useful to me, I study it, if not, I don't'', ''The main reason is represented by the existence of an exam that forces me to study''}.

Personal and emotional factors are related to each individual's passions:
\textit{''I am curious and I like to learn new things'', ''Because I like it'', ''If it is a topic that interests me''}. The personal relationship with the teacher seems to count less at this level, only 5.88\% of the students mentioning it, compared to 35.29\% of students mentioning \textit{''Interesting''} as a learning factor.

\textbf{What are the factors that discourage learning a topic or a course?}
The main discouraging factors are the mirror reflection of the most motivating factors: \textbf{Not Interesting} and \textbf{Not useful}" \textit{''It is hard, the teacher is demanding, exams are hard, I don't find it interesting, it wouldn't help me in the future'', ''It is useless in my development'', '' 
It does not captivates my attention'', ''the uselessness of course in life''}. Not liking the course and the topics or the teacher are discouraging factors: \textit{''Disinterest in the topic'', ''If I am not interested'', ''Because I do not like the teacher or the topics''}. \\

A large group mentioned \textbf{too much effort} to be a discouraging factor, others referred to stress, lack of time, theoretical and unuseful learning materials: \textit{''If there is too much to learn and I do not have the time necessary to retain everything'', ''Exam stress'', ''If it involves memorization and not thinking''}.

\begin{figure}[htbp!]
\centering
\includegraphics[width=0.45\textwidth,]{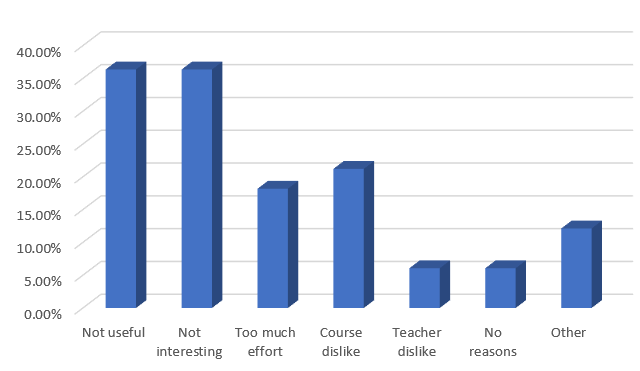}
\caption{Factors that discourage students to learn} \label{fig:ReasonsNotToLearn}
\end{figure}

Other answers reflect a learning mindset and a positive attitude: \textit{''There are no many reasons, it is always worth trying to learn something''}. The percentages for the factors that discourage learning are shown in Figure \ref{fig:ReasonsNotToLearn}.

\textbf{In conclusion:} Based on the answers to the questions, we can establish that the most important reasons are the how useful is the learned information and how interesting are the topics in the course. Personal relationship with the teacher seem to be less important, however, other factors such as too much effort or lack of time can negatively impact learning. 

\subsection{Q2 -  Which evaluation method is considered by students to be the most equitable and why?}

\begin{figure}[htbp!]
\centering
\includegraphics[width=0.450\textwidth,]{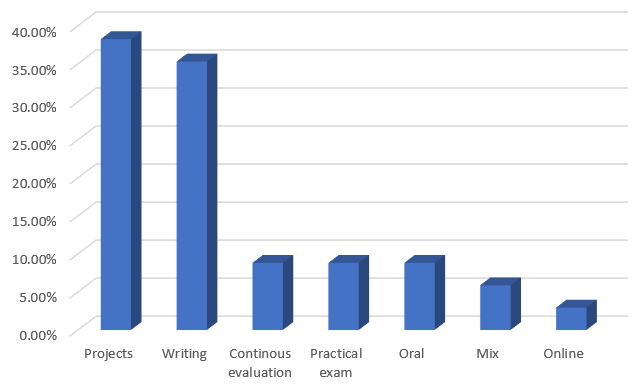}
\caption{Best forms of evaluations methods in student's opinions} \label{fig:bestEval}
\end{figure}

To find out, we decided to ask the students which forms of evaluation are in their opinion the best and which are the worst and why. Students prefer to be evaluated on written exams, practical projects, or practical examinations, or to have projects during the course. The percentages are shown in Figure \ref{fig:bestEval}.

\begin{figure}[htbp!]
\centering
\includegraphics[width=0.45\textwidth,]{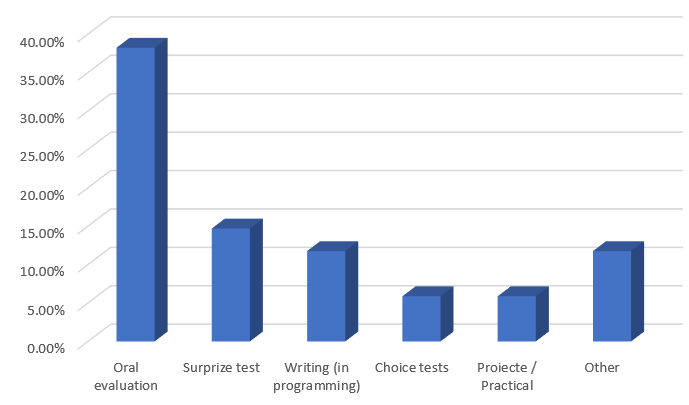}
\caption{Worst forms of evaluations methods in student's opinions} \label{fig:worstEval}
\end{figure}

As we did not want to create bias, we asked about the worst form of evaluation. If the oral evaluation scored low in the best forms of evaluation, it scored high in the worst form of evaluation. The surprise test was the second worst form of evaluation mentioned, as can be seen in Figure \ref{fig:worstEval}. Written examination was mentioned third as a bad form of evaluation, but only for programming courses: \textit{''Tests written on paper, especially for topics that are run in computer programs'', ''Reproducing, in writing, the theory or program sequence step-by-step, especially when a definition is formulated strangely and/or I do not have the possibility to test, in real time, what my program does''}

The students based their opinion on different reasons, the most important being \textbf{ stressful situations, not enough time} and learning \textbf{ memorizing / Unuseful information}.

\begin{figure}[htbp!]
\centering
\includegraphics[width=0.45\textwidth,]{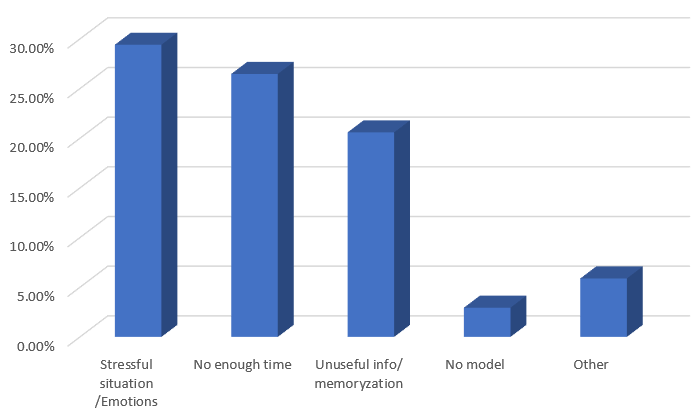}
\caption{Reasons for worst forms of evaluations methods in student's opinions} \label{fig:worstEvalReasons} 
\end{figure}

Students do not like stressful situations that appear during exams, especially in oral examinations; the stress is induced by the situation and the potential lack of time \textit{''Personally, I do not like oral listening'', ''The oral assessment because it is very stressful and you do not have much time to think'', '' Oral test because the student does not have enough time to think about the answers'', ''Practical assessment in a short time, because it does not assess logic but speed''}.
A large group mentions the theoretical examination, some students just mention it without additional explanations: \textit{''Theory exams''}, while others provide a more detailed reasoning: \textit{evaluation of theoretical knowledge}, \textit{because not all information is useful in practice, or used}, other claim that testing the assimilation of theoretical notions \textit{it is shifting the focus from the actual learning of the subject to some information that will be forgotten in a short time}.
The percentages can be visualized in Figure \ref{fig:worstEvalReasons}.

In conclusion, students prefer practical examination and written examination, but not when it involves writing programs on a piece of paper. Oral examination is considered by far the worst method of evaluation due to induced stress and lack of time. Also, a large group does not like theoretical exams that require information that is not useful in the long run. 

\subsection{Q3 - What do students think are the most effective ways to learn?}

\begin{figure}[htbp!]
\centering
\includegraphics[width=0.45\textwidth,]{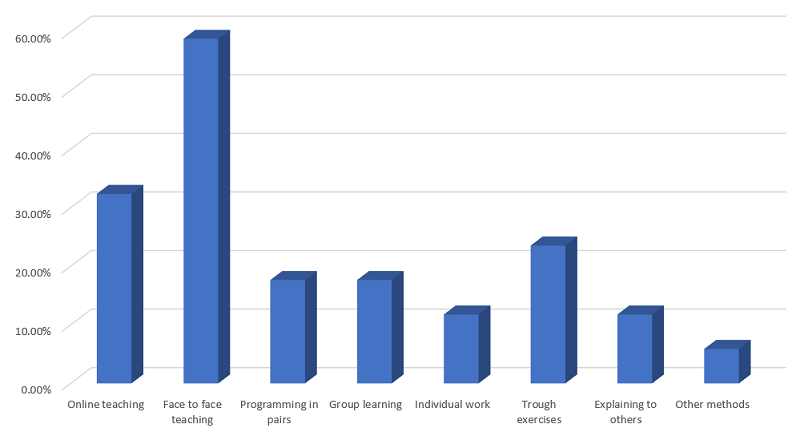}
\caption{Most effective learning methods in student's opinions} \label{fig:effectiveLearning} 
\end{figure}

This set of participants has special characteristics: students spend two years in high school in the COVID 19 pandemic, where most of the learning activities were online. Their exposure to online learning and courses has given them an extended experience over previous generations, so a high percentage, around 30\% of students mentioned it as an effective learning method. However, face-to-face was still the preferred learning method, scoring more than 50\% in student options. Programming in pairs, group learning, individual work, or practical problems also scored high in student's options as can be seen in Figure \ref{fig:effectiveLearning}. As a remark, some students mentioned more than one learning method: \textit{''Different online tutorials, face-to-face teaching and consulting the opinions of colleagues''}, so the sum of percentages exceeds 100\%.

Some students did not provide reasons for their learning preferences: \textit{''Different online tutorials, face-to-face teaching and consulting the opinions of colleagues'', ''face-to-face teaching, study groups'', ''Online courses and programming (learning) with a group of friends''}. Others provided a detailed reasoning for their choice: \textit{''Doing homework alone. Before doing my homework, I study the subject and then apply the learned things to the problems. I notice that the information settles very well in this way, not only in this course. Another good method of learning would be to teach a colleague / explain to someone who does not understand.''}
Learning by exercise is another method mentioned by students: \textit{''for me the best way of learning is practice. By doing many exercises, I learn much easier''}.

In conclusion, face-to-face learning is in the student's opinions the most effective learning method, followed by learning using exercises, group learning, and programming in pairs. So even for a generation that has extensive experience with online learning, they still prefer face-to-face learning,

\section{Treats to validity?}

Through the verification and application of community standards as described in \cite{ACM}, we attempted to reduce potential risks and addressed the validity threats mentioned in software engineering research \cite{ACM}. 
We considered the following possible threats: construct validity, internal validity, and external validity. 

The target participant set, participant selection, dropout contingency measures, and author biases were among the issues we focused on for internal validity.

\textbf{Selection and participant set.} 

We selected for the set of participants students who were enrolled in classes taught by one of the study's authors, chosen at random.

\textbf{Ethical concerns.} Before all students received the survey link, we informed them that participation was anonymous and voluntary. We also presented the scope of the research and how the collected data will be used.

\textbf{Rates of dropouts.} We had few options to increase participation in the study, as we were restricted by its voluntary nature. The only influence on the participation of the students in the study was to ensure that they understood the purpose of the survey. In addition, in an attempt to boost participation, we kept the survey open for two weeks and translated the questions into the students' native tongue. 

\textbf{Author Subjectivity.} We have taken measures to address the possibility of subjectivity in our data processing methodology. Following suggested data processing procedures, we as authors made sure that our work was cross-validated by examining each other's contributions.

\textbf{External validity.} We examine the possibility of extending and applying the study findings to the IT sector. The general research topic of the study is related to industry and the difficulties that various teams face when developing their human resources. We can conclude that the results of the study are useful to the local IT industry due to the topic.

\textbf{Validity of the construction.} We examined the surveys' questions for coherence, relevance, and pertinence. We developed the set of questions in three stages: first, a set of questions was proposed by the authors. All authors discussed and examined the questions to increase clarity. Only a subset of the questions that were most relevant to our study were chosen to be included in the survey (second step). In the final step, we verified the questions and decided which questions were required and which were optional.
After these procedures were completed, we created and distributed the survey to the participants. Since the data were gathered anonymously, the authors were unaware of the identities of the respondents.

\section{Discussions}

Upon examination of the data, we found that the students primarily attribute their learning to two things: the course's or the topics' level of interest and utility.

\textbf{Practical aspects.} There is a definite tendency to value and adhere to practical and passionate aspects. Students dislike mechanical learning or theoretical examinations where they are "asked to know information without use" or where they "will later forget." This tendency toward practical aspects is reinforced through answers to other questions. 
For example, their preference for projects or continuous evaluation are different aspects of the same disposition.

An intriguing feature about the students' view on evaluation is the absence of opinions regarding its fairness; preferences are solely based on stress, preferences, workload, and practicability. 

\textbf{Fear and stress.}
Numerous students indicated that the evaluation results were significantly impacted by their ability to control their emotions and stress. Students believe that oral evaluations are the worst due to stressful circumstances, time constraints, and teacher discussion — all of which make it difficult for them to handle. We have to ask ourselves after looking at these results \textit{Are these signs of a more emotional generation?} One more argument would be the fact that some of them acknowledge, admit, and express that they"are afraid". 
Stress and fear are strong emotions that can affect performance, attitude toward learning in general, and school / university specifically. 
Failure in a sphere of accomplishment can be a challenging experience that elicits feelings of guilt and fear, ranging from mild to severe conditions, such as atychiophobia, which can affect an individual's behavior and performance \cite{atkinson1957motivational}. Fear of being wrong, of being judged appears as a collateral reason for written examination: "because I can make mistakes without fear of being judged". 

Given that first-year students in the participant set had recently completed their high school education, these findings should raise concerns about the teaching methods and attitudes of high school teachers. 

\textbf{Reluctance to work.} 
Some students stated that they would prefer simpler courses that do not require a lot of work or learning time and effort when asked for the reasons why they do not learn.
Although there are certain factors that can make learning more difficult, such as scheduling classes during challenging times of the day, such as 6 to 8 p.m. or half past seven or eight a.m., they should not have a significant effect on learning. More future research can determine how these factors affect learning. However, stating that \textit{the time spent learning something new is too much} could be a reflection of a recent development: students' unwillingness to put in the necessary effort to learn.

Although the outcomes can serve as a guide for future modifications to instructional strategies, student preferences should not be the only factor considered. Their point of view should be regarded more as a guide for determining the best means of motivating and constraining them in order to optimize their academic performance. 

For instance, it's not always detrimental that a large portion of students exhibit a fear of failing. Because they consider learning hard and fear failure, it does not mean that we should simplify the learning content. Much better would be to search for methods to bias the natural tendency for anxious people to set either extremely low or very high aspirations in favor of the last, or to reduce the anxiety related to failure.

\section{Conclusions}

Our investigation of the students' primary motivations to acquire knowledge reveals that the most crucial factors are the perceived usefulness of the learned information and the level of interest in the course topics. Although personal relationships with teachers were found to be less influential, external factors, such as excessive effort or time constraints, can negatively impact the learning experience. These findings underscore the importance of aligning educational content with real-world applications and fostering student engagement to improve motivation.

Students believe that in-person instruction is the most effective way to learn, with exercise-based learning, group instruction, and pair programming coming in second and third.

Regarding the evaluation methods preferred by students, our research indicates a preference for practical and written examinations. However, it should be noted that the inclusion of writing programs on paper is met with resistance. Oral examinations emerged as the least favored method, attributed to induced stress and perceived time constraints. Additionally, a substantial portion of students expressed dissatisfaction with theoretical exams that they perceive as providing information that lacks practical utility in the long run. These insights highlight the need for educators to consider various assessment strategies that align with students' preferences and alleviate unnecessary stress.

Furthermore, our investigation of effective learning methods uncovered valuable information. The emphasis on practical and applicable knowledge is evident in students' preferences for hands-on learning experiences. Our findings underscore the importance of incorporating real-world scenarios and encouraging interactive learning approaches. As educators and institutions strive to optimize the learning process, understanding and integrating these preferences can contribute significantly to creating a more effective and engaging educational environment.

In summary, our research provides valuable information on the complex interplay of factors that influence students' motivations, preferences in evaluation methods, and effective learning strategies. By acknowledging and incorporating these findings, educators and educational institutions can foster a more conducive learning environment that reflects the diverse needs and preferences of students, ultimately enhancing the overall educational experience.


\balance

\bibliographystyle{apalike}
{\small
\bibliography{Example}}

\begin{thebibliography}{}

\bibitem[Araque et~al., 2009]{Araque}
Araque, F., Roldan, C., and Salguero, A. (2009).
\newblock Factors influencing university drop out rates.
\newblock {\em Computers \& Education}, 53:563--574.

\bibitem[Atkinson, 1957]{atkinson1957motivational}
Atkinson, J.~W. (1957).
\newblock Motivational determinants of risk-taking behavior.
\newblock {\em Psychological review}, 64(6p1):359.

\bibitem[Baquerizo et~al., 2020]{baquerizo2020motivacion}
Baquerizo, G. E.~B., M{\'a}rquez, F. A.~A., and Tobar, F. R.~L. (2020).
\newblock La motivaci{\'o}n en la ense{\~n}anza en l{\'\i}nea.
\newblock {\em Revista Conrado}, 16(75):316--321.

\bibitem[Braun et~al., 2019]{Braun19}
Braun, V., Clarke, V., Hayfield, N., and Terry, G. (2019).
\newblock {\em Thematic Analysis}, pages 843--860.
\newblock Springer Singapore.

\bibitem[Centre~for Higher Education Policy~Studies, 2015]{UEDropOut}
Centre~for Higher Education Policy~Studies, N. (2015).
\newblock Dropout and completion in higher education in europe.

\bibitem[de~Sousa et~al., 2022]{de2022characteristics}
de~Sousa, M.~M., de~Almeida, D. A.~R., Mansur-Alves, M., and Huziwara, E.~M. (2022).
\newblock Characteristics and effects of entrepreneurship education programs: a systematic review.
\newblock {\em Trends in Psychology}, pages 1--31.

\bibitem[George, 2020]{Marcus2020}
George, M.~L. (2020).
\newblock Effective teaching and examination strategies for undergraduate learning during covid-19 school restrictions.
\newblock {\em Journal of Educational Technology Systems}, 49(1):23--48.

\bibitem[Gutierrez-Aguilar et~al., 2023]{gutierrez2023learning}
Gutierrez-Aguilar, O., Talavera-Mendoza, F., Chica{\~n}a-Huanca, S., Cano-Villafuerte, S., and Sotillo-Vel{\'a}squez, J.~A. (2023).
\newblock E-learning and the factors that influence the fear of failing an academic year in the era of covid-19.
\newblock {\em Journal of Technology and Science Education}, 13(2):548--564.

\bibitem[Hazzan et~al., 2020]{Hazzan2020}
Hazzan, O., Lapidot, T., and Ragonis, N. (2020).
\newblock {\em Guide to teaching computer science}.
\newblock Springer.

\bibitem[Heublein et~al., 2017]{heuble2017}
Heublein, U., Ebert, J., Hutzsch, C., Isleib, S., K{\"o}nig, R., Richter, J., and Woisch, A. (2017).
\newblock Zwischen studienerwartung und studienwirklichkeit.
\newblock {\em Ursachen des Studienabbruchs, beruflicher Verbleib der Studienabbrecherinnen und Studienabbrecher und Entwicklung der Studienabbruchquote an deutschen Hochschulen}.

\bibitem[Lemay et~al., 2021]{Lemay2021TransitionTO}
Lemay, D.~J., Doleck, T., and Bazelais, P. (2021).
\newblock Transition to online teaching during the covid-19 pandemic.
\newblock {\em Interactive Learning Environments}, 31:2051 -- 2062.

\bibitem[Liu et~al., 2022]{liu2022increasing}
Liu, M., Gorgievski, M.~J., Qi, J., and Paas, F. (2022).
\newblock Increasing teaching effectiveness in entrepreneurship education: Course characteristics and student needs differences.
\newblock {\em Learning and Individual Differences}, 96:102147.

\bibitem[Martino, 1993]{Martino1993AGM}
Martino, L.~R. (1993).
\newblock A goal-setting model for young adolescent at-risk students.
\newblock {\em Middle School Journal}, 24:19--22.

\bibitem[Motogna et~al., 2021]{enase21}
Motogna, S., Suciu, D., and Molnar, A.-J. (2021).
\newblock Investigating student insight in software engineering team projects.
\newblock In {\em Proceedings of the 16th International Conference on Evaluation of Novel Approaches to Software Engineering - ENASE,}, pages 362--371. INSTICC, SciTePress.

\bibitem[Neuwirth et~al., 2021]{neuwirth2021reimagining}
Neuwirth, L.~S., Jovi{\'c}, S., and Mukherji, B.~R. (2021).
\newblock Reimagining higher education during and post-covid-19: Challenges and opportunities.
\newblock {\em Journal of Adult and Continuing Education}, 27(2):141--156.

\bibitem[Norem and Cantor, 1986]{Norem1986DefensivePH}
Norem, J.~K. and Cantor, N.~E. (1986).
\newblock Defensive pessimism: harnessing anxiety as motivation.
\newblock {\em Journal of personality and social psychology}, 51 6:1208--17.

\bibitem[Petrescu and Motogna, 2023]{petrescu2023perspective}
Petrescu, M. and Motogna, S. (2023).
\newblock A perspective from large vs small companies adoption of agile methodologies.
\newblock In {\em Proceedings of the 18th International Conference on Evaluation of Novel Approaches to Software Engineering - ENASE}, pages 265--272. INSTICC, SciTePress.

\bibitem[Petrescu et~al., 2023]{PETRESCU20231028}
Petrescu, M.-A., Pop, E.-L., and Tudor-{Dan Mihoc} (2023).
\newblock Students’ interest in knowledge acquisition in artificial intelligence.
\newblock {\em Procedia Computer Science}, 225:1028--1036.
\newblock 27th International Conference on Knowledge Based and Intelligent Information and Engineering Sytems (KES 2023).

\bibitem[{Ralph, Paul (ed.)}, 2021]{ACM}
{Ralph, Paul (ed.)} (2021).
\newblock {ACM Sigsoft Empirical Standards for Software Engineering Research}, version 0.2.0.

\bibitem[Salas, 2017]{salas2017teaching}
Salas, R.~P. (2017).
\newblock Teaching entrepreneurship in computer science: Lessons learned.
\newblock In {\em 2017 IEEE Frontiers in Education Conference (FIE)}, pages 1\--7. IEEE.

\bibitem[Seikkula-Leino et~al., 2010]{Seikkula}
Seikkula-Leino, J., Oikkonen, E., Ikävalko, M., Kolhinen, J., and Rytkola, T. (2010).
\newblock Promoting entrepreneurship education: The role of the teacher?
\newblock {\em Education + Training}, 52:117--127.

\bibitem[Sobral, 2020]{s2020}
Sobral, S. (2020).
\newblock Online teaching principles.
\newblock In {\em Erasmus Teaching Week (HMU)}.

\bibitem[Vegliante et~al., 2018]{v2018}
Vegliante, R., Miranda, S., and De~Angelis, M. (2018).
\newblock Online evaluation in the massive courses.
\newblock In {\em 11th annual International Conference of Education, Research and Innovation}, pages 8214--8220.

\bibitem[Wang et~al., 2021]{Wang21}
Wang, S., Bajwa, N.~P., Tong, R., and Kelly, H. (2021).
\newblock Transitioning to online teaching.
\newblock {\em Radical Solutions for Education in a Crisis Context: COVID-19 as an Opportunity for Global Learning}, pages 177--188.

\end{thebibliography}

\end{document}